\documentclass[10pt,aps,prd,reprint,nofootinbib,floats,floatfix,amsfonts,amssymb,amsmath,preprintnumbers,notitlepage,superscriptaddress]{revtex4-1}
\usepackage{aastex_macros}
\usepackage{float}
\usepackage{graphicx}
\usepackage[utf8]{inputenc}
\usepackage[british]{babel}
\usepackage{isomath}
\usepackage[protrusion=true,tracking=true,kerning=true,spacing=true,final,babel=true]{microtype}
\usepackage{physics}
\usepackage{xcolor}
\usepackage{url}
\usepackage[final]{hyperref}
\usepackage{nth}
\newcommand{\bs}[1]{\boldsymbol{#1}}

\newcommand{\odds}[2]{\mathcal{B}_{\rm #2}^{\rm #1}}

\begin{document}

\begin{flushleft}
KCL-PH-TH/2020-63 
\end{flushleft}
\begin{flushleft}
CERN-TH-2020-184 
\end{flushleft}

\title{Simultaneous estimation of astrophysical and cosmological stochastic gravitational-wave backgrounds with terrestrial detectors}
\author{Katarina~Martinovic}
\email{katarina.martinovic@kcl.ac.uk}
\affiliation{Theoretical Particle Physics and Cosmology Group, \, Physics \, Department, \\ King's College London, \, University \, of London, \, Strand, \, London \, WC2R \, 2LS, \, UK}
\author{Patrick~M.~Meyers}
\email{pat.meyers@unimelb.edu.au}

\affiliation{School of Physics, University of Melbourne, Parkville, VIC 3010, Australia}
\affiliation{OzGrav, University of Melbourne, Parkville, VIC 3010, Australia}

\author{Mairi~Sakellariadou}
\email{mairi.sakellariadou@kcl.ac.uk}
\affiliation{Theoretical Particle Physics and Cosmology Group, \, Physics \, Department, \\ King's College London, \, University \, of London, \, Strand, \, London \, WC2R \, 2LS, \, UK}
\affiliation{Theoretical Physics Department, CERN, Geneva, Switzerland}
\author{Nelson Christensen}
\email{nelson.christensen@oca.eu}
\affiliation{Artemis, Universit\'e C\^ote d'Azur, Observatoire de la C\^ote d'Azur, CNRS, Nice 06300, France}
\date{\today}    

\begin{abstract}
    The recent Advanced LIGO and Advanced Virgo joint observing runs have not claimed a stochastic gravitational-wave background detection, but one expects this to change as the sensitivity of the detectors improves. The challenge of claiming a true detection will be immediately succeeded by the difficulty of relating the signal to the sources that contribute to it.
    In this paper, we consider backgrounds that comprise compact binary coalescences and additional cosmological sources, and we set simultaneous upper limits on these backgrounds. We find that the Advanced LIGO, Advanced Virgo network, operating at design sensitivity, will not allow for separation of the sources we consider. Third generation detectors, sensitive to most individual compact binary mergers, can reduce the astrophysical signal via subtraction of individual sources, and potentially reveal a cosmological background. Our Bayesian analysis shows that, assuming a detector network containing Cosmic Explorer and Einstein Telescope and reasonable levels of individual source subtraction, we can detect cosmological signals $\Omega_{\rm{CS}} (25\,\rm{Hz})=4.5 \times 10^{-13}$ for cosmic strings, and $\Omega_{\rm BPL}(25\,\rm{Hz})= 2.2 \times 10^{-13}$ for a broken power law model of an early universe phase transition.
\end{abstract}{}

\maketitle

\section{Introduction}

A stochastic gravitational wave background (SGWB) is a random signal produced by many weak, independent and unresolved sources; it can be of a cosmological or astrophysical origin.
A variety of early universe processes, like quantum vacuum fluctuations during inflation, post-inflationary preheating, first order phase transitions, or topological defects (in particular cosmic strings) can lead to a SGWB \cite{Allen:1996vm,Maggiore:2000gv,Caprini:2018mtu}.

An astrophysical contribution to the SGWB comes from the superposition of unresolved gravitational-wave (GW) sources of stellar origin. This includes burst sources, like core collapse supernovae and the final stage of compact binary coalescences (CBCs), together with quasi-periodic long-lived sources like pulsars and the early inspiral phase of compact binaries \cite{Regimbau:2011rp, Christensen:2018iqi}. A detection of a SGWB can provide important astrophysical information about, for instance, the mass range for neutron star and black hole progenitors, or the rate of compact binaries \cite{Abbott:2017xzg}. It also sheds light on particle physics models beyond the Standard Model and the early stages of our universe. Advanced LIGO's second observing run saw no SGWB, and placed upper limits on frequency-independent and CBC backgrounds, as well as additional GW polarizations \cite{LIGOScientific:2019vic}.

Once a SGWB is successfully detected, there will be the challenge of identifying the sources that contribute to it. 
Untangling these signals will deepen our knowledge of merger rates and population models \cite{Boco:2019teq, Zhu:2012xw, Smith:2017vfk, Talbot:2018cva}, our understanding of exotic objects \cite{Blanco-Pillado:2017rnf, Ringeval:2017eww, Cardoso:2019rvt, Regimbau:2005ey} and in particular early universe models \cite{Abbott:2009ws, Guzzetti:2016mkm, Caprini:2018mtu, Hindmarsh:2017gnf}. 
Combining a stochastic analysis with information from individual events may provide the appropriate means to learn about GW sources \cite{Callister:2020arv}.

We have recently developed a parameter estimation analysis to distinguish between correlated magnetic noise that contaminates terrestrial GW detectors and a real GW signal~\cite{Meyers:2020qrb}. The low-frequency resonances in the Earth's global electromagnetic field could couple to the mass suspension system and electronics in the detectors, and mimic a SGWB. The method presented in \cite{Meyers:2020qrb} helps to minimize the possibility of a false detection. In this paper, we adapt that method, which is based on techniques already present in the literature~\cite{mandic:2012ulw,callister2017:owl,tsukada:2019olx}, to separate the astrophysical and cosmological contributions to the SGWB.
We first discuss second generation detectors like Advanced LIGO~\cite{TheLIGOScientific:2014jea} and Advanced Virgo~\cite{TheVirgo:2014hva}.
 We then move to third generation GW detectors, namely  Einstein Telescope (ET)~\cite{ET_Punturo} and Cosmic Explorer (CE)~\cite{Abbott_2017_3G,Reitze:2019dyk}, and comment on how the study could be adapted to the space-based LISA detector. The future detector networks require subtraction of the known compact binaries background from the stochastic signal, prior to the parameter estimation \cite{PhysRevLett.118.151105,Sachdev:2020bkk, PhysRevD.73.042001, PhysRevD.77.123010, PhysRevD.102.063009}, to ensure that the cosmological background is not obscured by the astrophysical one.

In Section \ref{sec:sources} we discuss the individual sources we choose for this study. We describe the GW signals injected and the analysis we perform  in Section \ref{sec:PE methods}. We summarize our results in Section \ref{sec:results} and make concluding remarks in Section~\ref{sec:discussion}.

\section{SGWB sources}
\label{sec:sources}

In this section, we discuss three different potential contributions to a SGWB. We discuss the astrophysical contribution from CBCs, and two cosmological sources of GWs. We consider cosmic strings, which are one-dimensional topological defects \cite{Vilenkin:2000jqa}, and an early-universe first order phase transition \cite{Witten:1984rs,Hogan:1986qda}.

\subsection{CBC background}
\label{subsec:cbc}

The CBC background is likely to be the largest contribution to the SGWB~\cite{Abbott:2017xzg}. Therefore, any attempt to measure other contributions to the background should be done in such a way as to simultaneously measure a CBC background and other contributions. The analytic model describing the CBC background depends on quantities such as redshift and merger rates~\cite{Regimbau:2011rp,Abbott:2017xzg}; the inspiral phase can be approximated as
\begin{equation}
\label{eq:omega_cbc_power_law}
   \Omega_{\rm CBC}(f) = \Omega_{2/3} \left(\frac{f}{25 \, \rm Hz} \right)^{2/3}.
\end{equation}
In the case of second-generation detectors, we can use this approximation freely~\cite{callister:2016nvl,Saffer:2020xsw}. When it comes to future GW detectors, however, the approximation cannot be applied to the entire frequency band. Instead, one must also include the contributions from the merger and ringdown phases that cause measurable deviations from this approximation~\cite{Sachdev:2020bkk,Saffer:2020xsw}. For the purpose of this study, we restrict ourselves to the range $(10 -100)$ Hz, the frequency range over which the approximation in Eq. (\ref{eq:omega_cbc_power_law}) is valid, even after individual source subtraction~\cite{Sachdev:2020bkk}.

The current estimate of the amplitude of the CBC spectrum from individual sources over the Advanced LIGO and Advanced Virgo frequency range is $\Omega_{\rm CBC} = 1.8^{+2.7}_{-1.7} \times 10^{-9}$, with 90\% confidence, at a reference frequency of 25~Hz~\cite{Abbott:2017xzg}. This estimate includes contributions from binary black holes, binary neutron stars, and black hole-neutron star systems. 

There are numerous studies on subtracting resolvable CBC signals from the data, and these can lead to a reduction in their contribution to the SGWB by as much as two orders of magnitude for binary black hole signals and one order of magnitude for binary neutron star signals~\cite{PhysRevLett.118.151105,Sachdev:2020bkk, PhysRevD.73.042001, PhysRevD.77.123010, PhysRevD.102.063009}. When considering future detectors like Einstein Telescope~\cite{ET_Punturo} and Cosmic Explorer~\cite{Abbott_2017_3G,Reitze:2019dyk}, we assume a scenario where such a subtraction has already been made -- following the results from~\cite{Sachdev:2020bkk}.


\subsection{Cosmic Strings}

A phase transition followed by a spontaneously broken symmetry can leave behind topological defects as remnants of a previous more symmetric phase. One particular class of such defects is cosmic strings (CS), line-like defects, generically formed within the context of grand unified theories \cite{Jeannerot:2003qv}. 

A network of cosmic strings is mainly characterized by the string tension $G\mu$, where $G$ is Newton's constant, and $\mu$ is the mass per unit length. The dynamics of a string network are driven by the formation of loops and the emission of bursts of GWs, predominantly from cusps and kinks. The superposition of these bursts leads to a SGWB over a large range of frequencies, making it a target for GW searches from pulsar timing arrays in the nHz band as well as the ground-based detectors we consider here \cite{Damour:2004kw, Lentati:2015qwp, Siemens:2006yp}. 

In the high-frequency regime we consider, $(10-100)$ Hz, the spectrum of the SGWB is flat, i.e. $\Omega_{\rm CS} (f) = \rm const$ \cite{Gouttenoire:2019kij}, and it only depends on the averaged total power emitted by a loop, and the total number of loops. A SGWB analysis can thus put a limit on the string tension, and consequently on the energy scale of the phase transition leading to the formation of these objects.

The 95\% credible upper limit placed after the first two LIGO observation runs, assuming a uniform prior, is $\Omega_{\rm CS}=6.0 \times 10^{-8}$ \cite{LIGOScientific:2019vic}. This implies upper bounds to the string tension, 
$G\mu\leq 1.1 \times 10^{-6}$ and $G\mu\leq 2.1 \times 10^{-14}$,
for the loop distribution models
\cite{Blanco-Pillado:2013qja} and \cite{Lorenz:2010sm},
respectively.

\subsection{First Order Phase Transitions}
\label{subsec:PT}

If a phase transition occurred at temperatures $(10^4 - 10^5)~\rm TeV$, the corresponding GW spectrum would be observed in the $(10-100)$ Hz frequency range we consider \cite{vonHarling:2019gme}. The phase transition associated with the breaking of Peccei-Quinn symmetry, for instance,  could have happened at such high temperatures,
leading to the QCD axion, a well motivated extension to the Standard Model. In this scenario, the growth of the true vacuum bubbles, and their subsequent collisions, give out GWs due to several effects \cite{Caprini_2016}. The strongest of those is most likely due to sound waves from bubble growth in plasma. The turbulence of the plasma in which the bubbles grow can also produce GWs. Finally, GWs are emitted due to collision of the scalar wall profiles.
There exist numerical \cite{Huber:2008hg, Hindmarsh:2013xza, Hindmarsh:2015qta} and analytic \cite{Caprini:2007xq, Jinno:2016vai} models for the shape of $\Omega_{\rm GW}$ as a function of frequency for each of these contributions. 

The frequency spectrum of the SGWB produced by most models can be captured by a smoothed broken power law:
\begin{equation}
  \Omega_{\rm BPL} =
              \Omega_*~\Big(\frac{f}{f_*}\Big)^{\alpha_1}~ \Bigg[1+\Big(\frac{f}{f_*}\Big)^{\Delta}\Bigg]^{(\alpha_2-\alpha_1)/\Delta}.
  \label{eq:bpl}              
\end{equation}
For example, numerical simulations find the GW spectrum due to the sound waves in the plasma \cite{Weir:2017wfa}

\begin{equation}
     h^2\Omega_{\rm SW}(f) =  F(\beta, H_*, \kappa_{\rm sw}, \alpha, g_*, v_w) ~\frac{(f/f_{\rm sw})^3}{[1+0.75(f/f_{\rm sw})^2]^{7/2}},
     \label{eq:omega_sound_waves}
\end{equation}
where $\beta$ is the transition strength,  $H_*$ is the Hubble constant at the time of GW production, $\kappa_{\rm sw}$ is the efficiency factor, $\alpha$ is the ratio of latent heat released in the phase transition to the heat of the radiation bath, $g_*$ is the number of relativistic degrees of freedom, $v_w$ is the bubble wall velocity, and $f_{\rm sw} = f_{\rm sw}(\beta, H_*)$ is the peak frequency.

If we use Eq. (\ref{eq:bpl}) to approximate Eq. (\ref{eq:omega_sound_waves}), then we have $\alpha_1=3$, $\alpha_2=-4$ and $\Delta=2$. Relating $\Omega_*$ and $f_*$ to the long list of physical parameters that control the phase transition is beyond the scope of this study.

\section{Model selection and parameter estimation}
\label{sec:PE methods}

We undertake a Bayesian parameter estimation and model selection study. For a single GW detector pair, $ij$, the log-likelihood is
\begin{align}
    \log p(\hat C_{ij}(f) | \boldsymbol{\theta}_{\rm GW})=&-\frac{1}{2}\sum_{f}\frac{\left[\hat C_{ij}(f) - \Omega_{\rm GW}(f, \boldsymbol{\theta}_{\rm GW}) \right]^2}{\sigma_{ij}^2(f)} \nonumber\\&-\frac{1}{2}\sum_{f} \log\left[2\pi\sigma_{ij}^2(f)\right],
    \label{eq:pe_ms:likelihood}
\end{align}
where $\Omega_{\mathrm GW}(f)$ is the model spectrum and $\boldsymbol{\theta}_{\rm GW}$ are the parameters that define the model. The cross-correlation estimator, $\hat C_{ij}(f)$, is calculated from detector data and is discussed in detail in \cite{Romano:2016dpx,LIGOScientific:2019vic,Meyers:2020qrb}. We extend this analysis to include three GW detectors by adding log-likelihoods for the individual pairs to construct a multiple-baseline log-likelihood.

To compare two models, $\mathcal{M}_1$ and $\mathcal{M}_2$, and make statements about which is more favourable by the data, we utilise Bayes factors,
\begin{align}
\odds{\mathcal{M}_1}{\mathcal{M}_2} = \frac{\int \textrm{d}\bs\theta~ p(\hat C_{ij}(f) | \bs\theta, \mathcal{M}_1)p(\bs\theta|\mathcal{M}_1)}{\int \textrm{d}\bs\theta~ p(\hat C_{ij}(f) | \bs\theta, \mathcal{M}_2)p(\bs\theta|\mathcal{M}_2)}
\label{eq:pe_ms:bayes_factor}
\end{align}
where $p(\bs\theta|\cdot)$ is the prior probability of our parameters given a choice of model. The integrand in Eq.~(\ref{eq:pe_ms:bayes_factor}) is the joint posterior distribution of the model parameters, which is evaluated as part of the evaluation of the Bayes factors.

For large and positive values of $\ln\odds{\mathcal{M}_1}{\mathcal{M}_2}$, there is strong evidence for $\mathcal{M}_1$ over $\mathcal{M}_2$. Likewise, large and negative values show preference for $\mathcal{M}_2$. Relating this quantity to a frequentist SNR statistic~\cite{Allen:1996vm}, we have $\rm ln \, \mathcal{B} \propto \rm SNR^2$ \cite{Romano:2016dpx}.
We use the nested sampler \texttt
{dynesty} through the front-end package {\sc Bilby} to evaluate Bayes factors for our models, as well as posterior distributions on the parameters.

While the posterior distribution of $\bs\theta_{\rm GW}$ is evaluated in conjunction with Bayes factors, we can also analytically calculate a bound on covariance between model parameters using the information matrix. This is has been used for estimating parameter covariance for SGWB models in other studies as well~\cite{Parida:2015fma,Kuroyanagi:2018hcz,Saffer:2020xsw}. For the case of a Gaussian likelihood with uncorrelated measurements (frequency bins) with an unbiased estimator, the information matrix is given by
\begin{align}
\mathcal I_{ij}(\bm\theta) = \sum_f\sigma(f)^{-2}\left(\frac{\partial\Omega_{\rm GW}(f,\bm\theta)}{\partial\theta_i}\right)\left(\frac{\partial\Omega_{\rm GW}(f,\bm\theta)}{\partial\theta_j}\right).
\label{eq:pe_ms:information matrix}
\end{align}
The covariance between model parameters is theoretically bounded below by the inverse of the information matrix
\begin{align}
\textrm{cov}_{\bm\theta}\left(\theta_{i}, \theta_{j}\right) \geq \left[\mathcal I^{-1}(\bm\theta)\right]_{ij}.
\label{eq:pe_ms:covariance_bound}
\end{align}
This bound, known as the Cramér-Rao lower bound, can be exceeded by including, e.g. informative prior information. However, the structure of the information matrix can still offer valuable insight into the degeneracy of certain model parameters with one another and offer an intuitive picture of the parameter estimation problem.

\subsection*{Injected Signal}
\label{sec:injections}

We consider two types of injections: one containing a CBC and a cosmic strings background, and another one containing a CBC and a background due to phase transitions, see Table \ref{tab:injections}. The background labelled here as CBC refers to what is left once we subtract the known CBC contribution, i.e. it is the unresolved astrophysical background. For the second injection, we choose a broken power law with exponents $\alpha_1=3$, $\alpha_2=-4$, and $\Delta=2$ which best describes $\Omega_{\rm SW}$, the sound wave contribution to $\Omega_{\rm GW}$. 
In this case our Bayesian search estimates the peak frequency, $f_*$, as well as the amplitude of the smooth broken power law, $\Omega_*$.

The injection strengths we choose vary from one detector network to another. 
The instrumental noise is included at the level of the design sensitivity curves of the detectors. We consider O4 sensitivity for Advanced LIGO and Advanced Virgo \cite{Aasi:2013wya}, ET-D for the Einstein Telescope \cite{Hild:2010id} and CE Wideband for the Cosmic Explorer \cite{Evans:2016mbw}. The same prior is used for the recovered amplitudes, $\Omega_{2/3}, \Omega_{\rm CS}, \Omega_*$, all of them log uniformly distributed between $10^{-15}$ and $10^{-8}$. All results are presented for 1 year observation time.

\begin{table}[H]
    \centering
    \begin{tabular}{lccccccc}
    \hline
       & $\Omega_{\rm GW}(f)$  & GW parameters, $\boldsymbol{\theta}_{\rm GW}$\\
        \hline
    Injection 1   & $\Omega_{\rm CBC}(f) + \Omega_{\rm CS}(f)$ & $(\Omega_{2/3}, \Omega_{\rm CS})$ \\
    Injection 2 & $\Omega_{\rm CBC}(f) + \Omega_{\rm BPL}(f)$ & $(\Omega_{2/3}, \Omega_*,f_*)$  \\
        \hline
    \end{tabular}
    \caption{GW spectra injected, and the parameters estimated in the analysis.}
    \label{tab:injections}
\end{table}

\section{Results}
\label{sec:results}

We present results on source separation for a SGWB detection with different sets of GW detector networks.

\subsection{Advanced LIGO and Advanced Virgo}

In this section we consider separation of a CS signal from a CBC signal with the current detector network operating at design sensitivity. We vary injection strengths, with 25 injections log uniformly distributed between $\Omega_{2/3} \in (10^{-9.4}, 10^{-8.4})$. These values were chosen by using 90\% limits on CBC background from \cite{Abbott:2017xzg}. We explore the flat cosmic strings spectrum with 35 injections log-uniformly distributed between $\Omega_{\rm CS} \in (10^{-9.4}, 10^{-7.4})$. The upper limit of the injection range is consistent with constraints placed on a cosmic strings SGWB spectrum from data in the first two observational runs \cite{LIGOScientific:2019vic}. The Bayes factors we find are too low to differentiate between the signals,  indicating that one cannot distinguish models that include both spectra from models that include only a CBC background. Other methods, which seek to model the contribution from individual CBCs on shorter time-scales, along with an isotropic, flat background propose ways of overcoming these obstacles~\cite{Biscoveanu:2020gds, Mukherjee:2019oma}.

\subsection{Third Generation Detectors}

\begin{figure}[H]
          \includegraphics[width=0.45\textwidth]{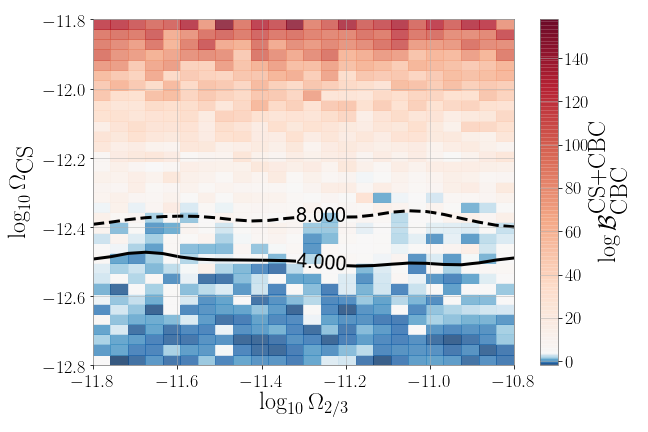}
          \includegraphics[width=0.45\textwidth]{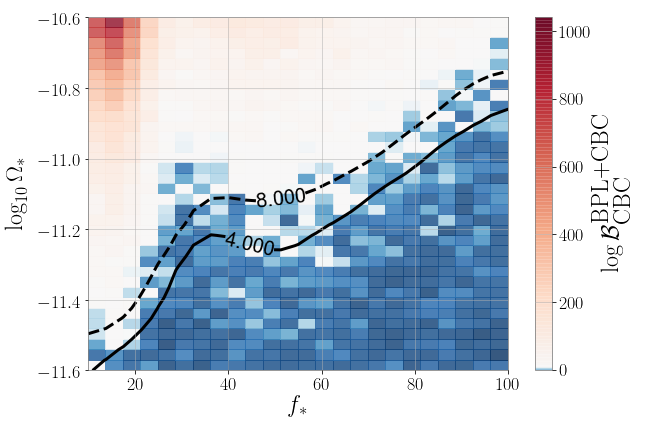}
    \caption{Variation of log Bayes factor with the injected power laws for cosmic strings (top panel), and a  first order phase transition (bottom panel).
    The residual CBC amplitude for the bottom panel is $\Omega_{2/3}=1\times10^{-11}$ for all of the injections.
    The contour plots show values of $\log \cal{B}=$ 4 or $\log\cal{B}=$ 8, which is roughly when we start to see significant bias to one of the models, since $\log\cal{B}=$ 8 corresponds to approximately SNR = 4.}
    \label{fig:ce_et}
\end{figure}

Operating at their design sensitivity, the Advanced LIGO, Advanced Virgo network cannot achieve source separation of a detected stochastic signal. We therefore pursue studies in future detectors. As was done in \cite{Sachdev:2020bkk}, we consider a network of Cosmic Explorer detectors at the Hanford and Livingston locations, and Einstein Telescope at the Virgo site. Fig.~1 in \cite{Sachdev:2020bkk} estimates that after individual source subtraction, the residual CBC contribution to the SGWB is dominated by unresolved binary neutron star mergers at the level of $\sim 10^{-11}$ at 10 Hz. We therefore use a log-uniformly distributed range of $\Omega_{2/3} \in (10^{-11.8}, 10^{-10.8})$ at 25 Hz in the top panel of Fig. \ref{fig:ce_et}, and fix $\Omega_{2/3} = 1 \times 10^{-11}$ for all the injections in the bottom panel. We then use comparable signal strengths for the cosmological contributions, in particular $\Omega_{\rm CS} \in (10^{-12.8},10^{-11.8})$ and $\Omega_{*} \in (10^{-11.6}, 10^{-10.6})$.

The GW selection effect could favor the detection of  the best oriented and located sources, especially  at larger redshift, disqualifying the assumption of an isotropic SGWB in the standard cross-correlation statistic. This leads to a systematic bias in the residual background and hence to a correction for the overlap reduction function \cite{Meacher:2014aca,Meacher:2015rex}. This could provide another way for discriminating between an astrophysical from a cosmological background which we will investigate in a future study.

From the top panel of Fig. \ref{fig:ce_et}, we see that we start to confidently separate a flat spectrum from the residual CBC signal for $\Omega_{\rm{CS}}=4.5 \times 10^{-13}$ . Cosmic strings backgrounds lower than this get lost in the unrecovered CBCs and cannot be singled out. Our sensitivity allows constraints to be placed on the string tension as low as $G\mu \leq 3.0 \times 10^{-17}$ and $G\mu \leq 4.0 \times 10^{-19}$, for the cosmic string loop distribution models \cite{Blanco-Pillado:2013qja} and \cite{Lorenz:2010sm},
respectively. Similar sensitivity to a cosmic strings spectrum is expected from the space-based LISA detector, whereas the Square Kilometer Array is expected to at most probe $G\mu$ values 3 or 4 orders of magnitude less sensitive \cite{Auclair:2019wcv}. 

As for a broken power law background due to an early universe phase transition, we find that the Cosmic Explorer and Einstein Telescope network's sensitivity is highly dependent on the break frequency of the spectrum, see bottom panel of Fig.~\ref{fig:ce_et}. The most conservative estimate we find of a detectable BPL signal (i.e. with $\log\cal{B}=$ 8), is the one associated with $f_*=100$~Hz, $\Omega_*= 1.8 \times 10^{-11}$. Taking into account injected values for $\alpha_1, \alpha_2, \Delta, f_*$, we estimate a stochastic signal of amplitude $\Omega_{\rm BPL}= 2.2 \times 10^{-13}$ at 25 Hz.

\begin{figure}[H]
    \centering
    \includegraphics[width=0.45\textwidth]{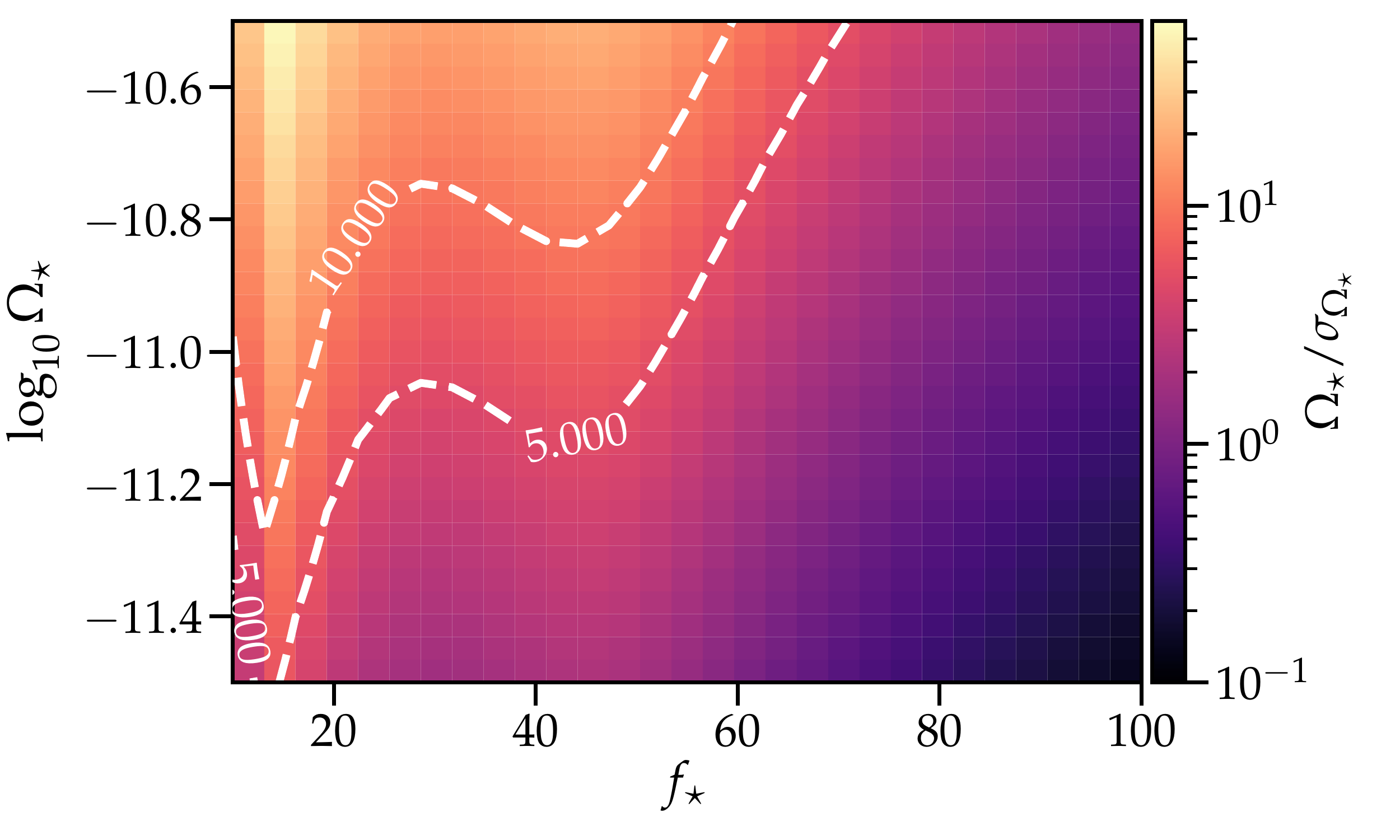}
    \includegraphics[width=0.45\textwidth]{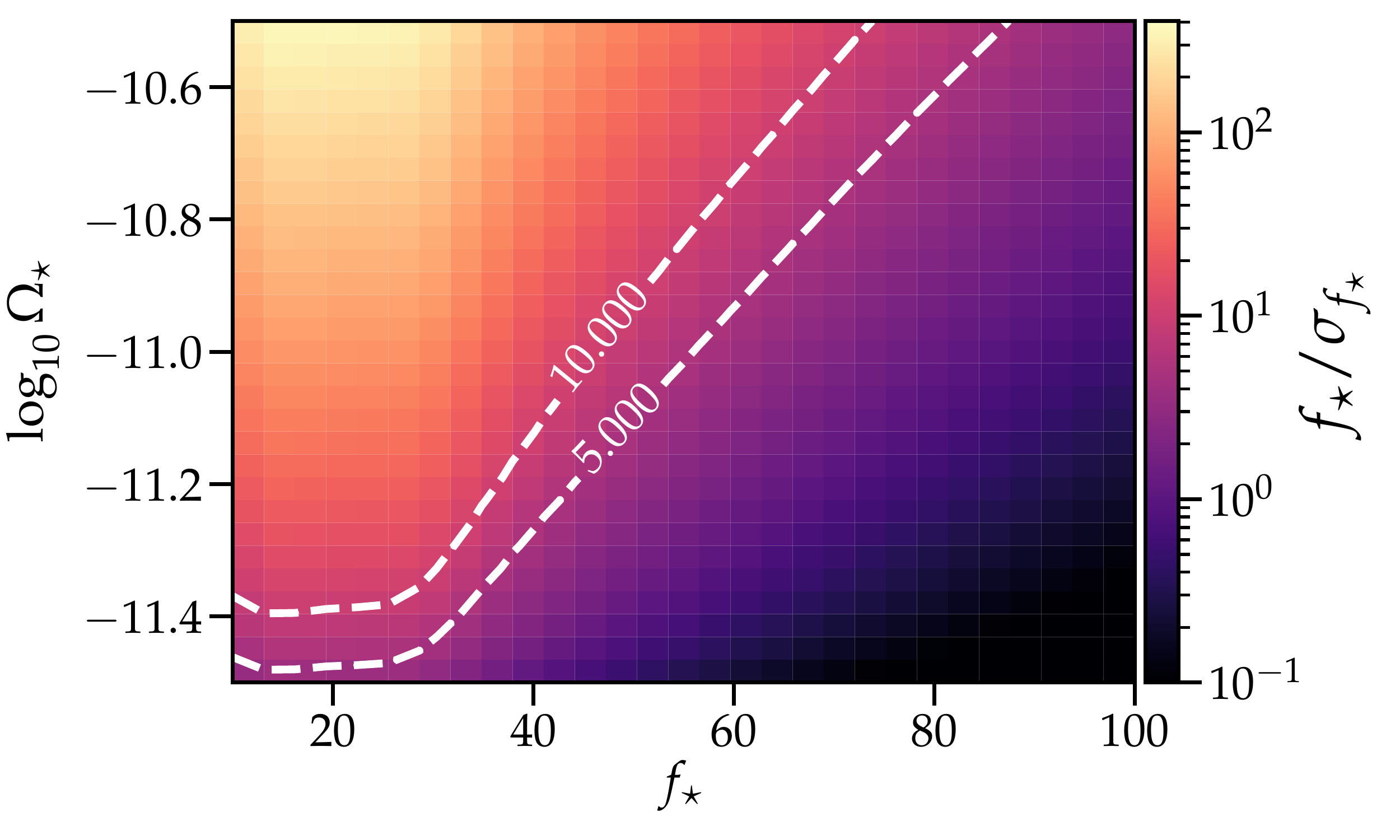}
    \includegraphics[width=0.45\textwidth]{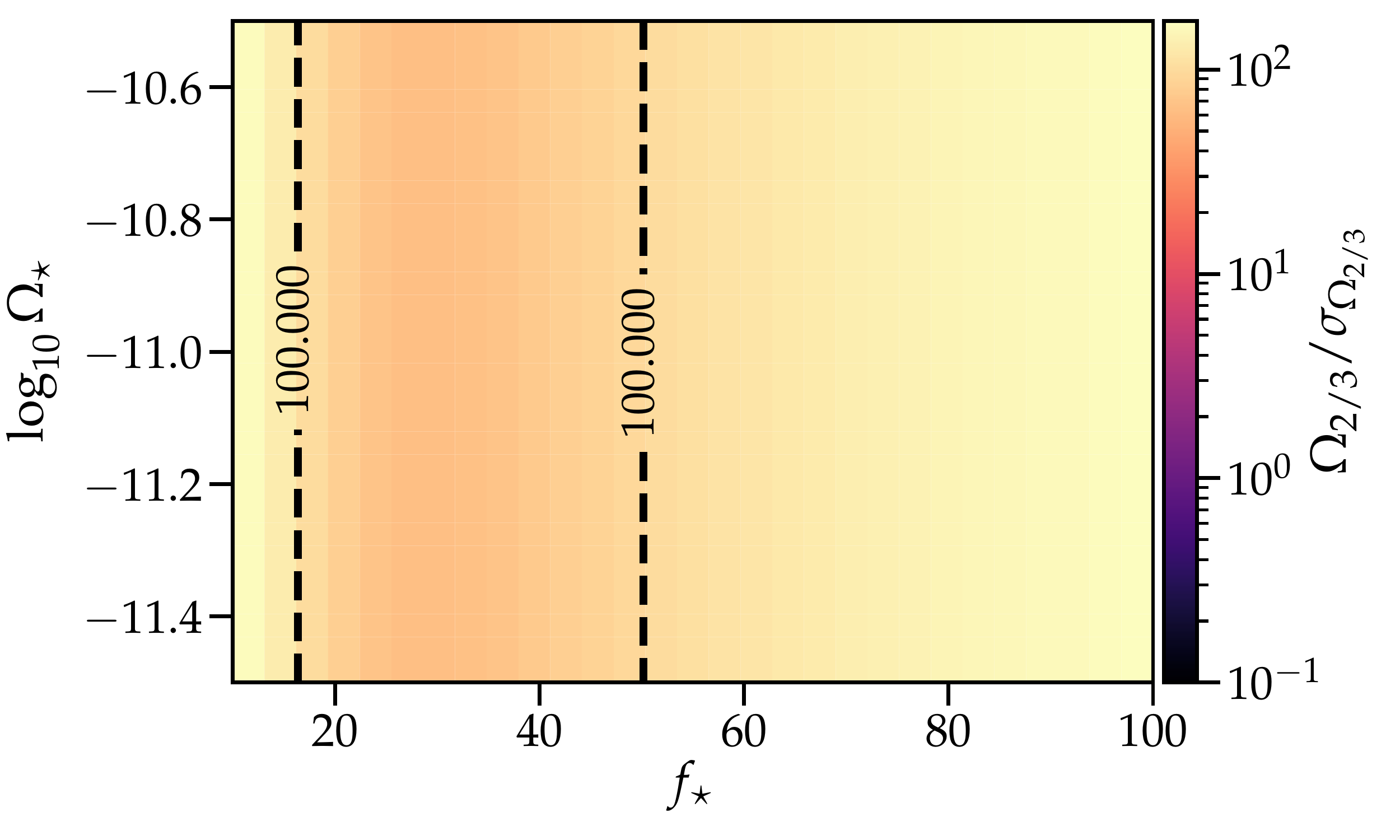}
    \caption{Precision with which we can measure $f_*$, $\Omega_*$, and $\Omega_{2/3}$ for the broken power law model, where $\sigma$ for each parameter is estimated using the bound in Eq. (\ref{eq:pe_ms:covariance_bound}). The model parameter used for $f_*$ and $\Omega_*$ is given by the value of the $x$- and $y$-axes respectively. The residual CBC injection is  $\Omega_{\rm 2/3}$ = $1 \times 10^{-11}$ for all simulations.}
    \label{fig:results:cr_bound_plots}
\end{figure}

We also look at the precision with which we can measure $\Omega_*, f_*$ and $\Omega_{2/3}$ using the covariance bound in Eq. (\ref{eq:pe_ms:covariance_bound}). We use $f_*/\sigma_{f_*}$ as a proxy for the precision of our $f_*$ measurement, with $\sigma_{f_*} = \left[\mathrm{cov}(f_*,f_*)\right]^{1/2}$ estimated from Eq. (\ref{eq:pe_ms:covariance_bound}) (and analogous expressions for $\Omega_*$ and $\Omega_{2/3}$). In Fig.~\ref{fig:results:cr_bound_plots}, we show the theoretical bound on this precision for $\Omega_*,f_*,\Omega_{2/3}$ as a function of the strength and shape of the broken power law background. In all three panels, the horizontal axis is $f_*$ and the vertical axis is $\log_{10}\Omega_*$. The color is the precision statistic discussed above. For all three panels, we have fixed $\Omega_{2/3}=1 \times 10^{-11}$. The broken power law model parameters are best estimated when $f_* \approx 20~\rm{Hz}$ and are improved as $\Omega_*$ increases. Interestingly, the theoretical precision with which we measure $\Omega_{\rm 2/3}$ is independent of $\Omega_{*}$, but is dependent upon the shape, which is governed by $f_*$. This is because elements of the information matrix in Eq.~(\ref{eq:pe_ms:information matrix}) that involve derivatives of $\Omega_*$ and $\Omega_{2/3}$ are independent of both $\Omega_*$ and $\Omega_{2/3}$ because these variables appear linearly in separate terms of the combined model $\Omega_{\rm BPL}(f) + \Omega_{\rm CBC}(f)$. This means that the variance of $\Omega_*$ and $\Omega_{2/3}$, and the covariance between them are independent of these amplitudes. These variables are still correlated with each other -- merely the covariance between them is independent of the values themselves.

\section{Discussion}
\label{sec:discussion}

We have looked at current, and future, terrestrial GW detectors to see if we can successfully perform source separation of a detected SGWB signal. This is an important task, since it allows us to relate a detection to physical theories underlying it and perhaps give us a hint of beyond Standard Model physics. Although Advanced LIGO and Advanced Virgo sensitivity is not sufficient to separate sources, we find promising results for the third generation of detectors such as the Einstein Telescope~\cite{ET_Punturo} and Cosmic Explorer~\cite{Abbott_2017_3G,Reitze:2019dyk}.

Our study concerns the frequency range for ground based detectors. However, our methods will certainly be applicable for the future space based detector, LISA~\cite{2017arXiv170200786A}. The LISA observational band offers an exciting possibility to observe GWs from phase transitions~\cite{Caprini_2016}. Much work has been done to develop methods to characterize an arbitrary SGWB spectrum~\cite{Caprini:2019pxz}, as well as techniques to distinguish a cosmologically produced SGWB from galactic binaries~\cite{Adams_2014}, a binary black hole produced background~\cite{Biscoveanu:2020gds}, and instrumental noise~\cite{Adams_2010,Adams_2014}.
A similar spectral separation study for LISA would be more complicated due to the nature of the time delay interferometry~\cite{Tinto_2014} and the necessity to simultaneously estimate the LISA noise. As such, we will apply the methods we have developed to LISA in a future study.

This analysis can be additionally extended by considering other cosmological sources of a SGWB. One can for instance consider the minimal Pre-Big-Bang model for which $\Omega_{\rm GW} (f)$ today scales as $f^3$ at the low frequency end of the spectrum, whereas  in the high frequency range its behaviour depends on a dimensionless free parameter of the model \cite{PhysRevD.47.1519}. Furthermore, one can consider the full analytical model for a CBC background, thereby expanding the studied frequency range.

Let us note that we have not considered Schumann noise, which could contaminate the stochastic background leading to a false detection \cite{Thrane:2013npa}. This issue may be of concern for
LIGO/Virgo, but as discussed above, there can be no source separation for current detectors.
Einstein Telescope is expected to have weaker coupling to Schumann noise due to heavier test masses, and the predicted magnetic budget is well below the sensitivity curves for post-Wiener filtering \cite{Amann:2020jgo}. Investigation of the magnetic budget for the Cosmic Explorer has not been undertaken, and there is uncertainty over what the magnetic contamination will look like. As studies on third generation detectors advance, we plan to extend our work with a detailed treatment of correlated magnetic noise.




\section{Acknowledgements}
The authors would like to thank the LIGO/Virgo Stochastic Background group for helpful comments and discussions. In particular Alexander Jenkins for his numerical simulations of cosmic strings spectra, and Tania Regimbau for useful discussions. The authors are grateful for computational resources provided by the LIGO Laboratory and supported by National Science Foundation Grants PHY-0757058 and PHY-0823459.

 Parts of this research were conducted by the Australian Research Council Centre of Excellence for Gravitational Wave Discovery (OzGrav), through project number CE170100004.
 K.M. is supported by King's College London through a Postgraduate International Scholarship. M.S. is supported in part by the Science and Technology Facility Council (STFC), United Kingdom, under the research grant ST/P000258/1.
 N.C. acknowledges support from National Science Foundation grant PHY-1806990.  
 
 This paper has been given LIGO DCC number P2000470.

 Numerous software packages were used in this paper. These include \texttt{matplotlib}~\cite{Hunter:2007}, \texttt{numpy}~\cite{numpy}, \texttt{scipy}~\cite{2020SciPy-NMeth}, \texttt{bilby}~\cite{Ashton:2018jfp}, \texttt{cpnest}~\cite{cpnest}, \texttt{ChainConsumer}~\cite{Hinton2016}, \texttt{seaborn}~\cite{michael_waskom_2014_12710}.

\newpage
\let\cleardoublepage\clearpage
\bibliographystyle{unsrt}
\bibliography{sgwb}

\end{document}